# Influence on extracted complex refractive index from phase inaccuracy of reflection-type THz-time-domain spectroscopy

Jeong Woo Han[1]

[1]Department of Physics Education, Chonnam National University, Gwangju 61186, South Korea
*jwhan@chonnam.edu

**Abstract:** Reflection-type terahertz (THz) time-domain spectroscopy (THz-TDS) enables the measurement of optical properties of opaque samples in the THz frequency range, e.g., carrier density and mobility. In this study, we examine the influence of phase inaccuracy on the extracted complex refractive index from reflection-type THz-TDS. Phase inaccuracy often arises from the placement mismatch between the perfect reflector, serving as the reference, and the target samples in reflection-type THz-TDS. By considering two representative systems, where free and bound carriers dominate optical properties, and introducing arbitrarily shifted placement mismatch, we confirm that significant errors in the extracted complex refractive index occur when the mismatch position exceeds 3 μm.

## 1. INTRODUCTION

Terahertz (THz) time-domain spectroscopy (THz-TDS) enables the investigation of the light-matter interaction in the THz frequency range (1 THz ~ 4 meV) quantitatively, by giving us the complex refractive index $\tilde{n}=n+ik$.[1-4] Therefore, THz-TDS has been utilized across various research and industrial fields, such as the investigation of vibrational modes in molecules[5-9] and carrier dynamics in semiconducting materials[10-12], which exhibit signatures in the THz frequency range. Because the measurement result of THz-TDS is recorded in the time domain, two-independent quantities, i.e., amplitude and phase, can be obtained in the frequency domain by applying the Fourier transform. This fact leads to the fact that $n$ and $k$ values can be extracted by the experiment data of THz-TDS with a single shot measurement, contrasting to other spectroscopic techniques.[13] For example, Fourier-transform infrared spectroscopy (FT-IR) is usually combined with the UV-visible ellipsometry to obtain the broad range of reflection and/or transmission spectrum and needs the mathematical treatment (Kramer-Kronig relation) to converge the experimental data into $\tilde{n}$.[14-16]

The experimental result of THz-TDS is recorded in the time domain, making THz-TDS phase sensitive measurement technique.[17-19] The phase information of the THz pulse is determined by its trajectory.[20,21] In order to extract reliable $\tilde{n}$, the accurate measurement of the phase information of the sample and the reference is necessary. It means that the phase information of the sample should be determined by the sample itself, not by any artificial geometrical factors. In this context, the sample placement plays a minor role in the measurement of the phase information in the transmission geometry because the reference spectrum is commonly obtained without the sample. In reflection geometry, the corresponding reference signal is usually obtained by replacing the sample with a known reflector in which the noble metals showing almost 100 % reflectance in the THz frequency region are usually employed. Ensuring identical geometrical information between the sample and the reference reflector during the replacement procedure is very challenging, meaning that the artificial geometrical information can be easily involved in the phase information. This challenge triggers many studies in the development of the new type of THz-TDS in reflection geometry, which allows us to obtain reliable $\tilde{n}$. Most of the proposed methodologies focus on eliminating the error source, i.e., the replacement process, which can be achieved by employing the mechanical conjugation between the target sample and the reference.[20-22] By employing this conjugated structure, the sample and the reference THz signal can be obtained by a single measurement. However, this proposed technique has the risk of damaging the sample during the process for the mechanical conjugation, especially when the sample is fragile. Therefore, there are still high demands for utilizing conventional reflection-type THz-TDS with an improved replacement technique. In this aspect, the quantitative error analysis on $\tilde{n}$ is required in the presence of the uncertainty of the placement between the sample and the reference.

In this study, we investigate the accuracy of extracted $\tilde{n}$ from reflection type THz-TDS under the phase imprecision resulting from the placement mismatch between the sample and the reference. We artificially adjust the position of the THz pulse recorded in the time domain to reproduce the placement mismatch thereof. Two representative systems are considered, which are metallic and insulating systems. From our quantitative analysis, the inaccuracy of $\tilde{n}$ increases drastically when the placement mismatch exceeds 3 µm.

## 2. RESULTS AND DISCUSSION

Figure 1 illustrates the geometry of reflection type THz-TDS. For a systematic analysis of the errors on the extracted $\tilde{n}$ due to the placement mismatch D between the sample and the reference, the signals of incident $E_{inc}$ and reflected $E_{rec}$ THz are required as a function of D. To obtain $E_{inc}$, we measure the THz pulse emitted from a commercialized photo-conductive antenna (PCA) combined with the pulsed laser system. Note that we used the same type of PCA for the detection of the THz pulse. By using the known $\tilde{n}=n+ik$ combined with obtained $E_{inc}$, we can simulate $E_{ref}$ under perfect alignment conditions, as:

$$E_{rec}(\omega)=IF\left(F(E_{Inc}(\omega))\cdot\frac{1-n(\omega)-ik(\omega)}{1+n(\omega)+ik(\omega)}\right) \quad (1)$$

where $F$ and $IF$ are the Fourier and inverse Fourier transforms, respectively. $\omega$ denotes frequency. We arbitrarily adjust the THz pulse position in the time domain to account for D. Note that the adjusted time delay $\Delta t$ can be calculated by using the simple relation $\Delta t=D/c$, where $c$ is the speed of light. For example, D of 300 µm corresponds to 1 ps in the time domain. By following this procedure described above, we can obtain the THz pulse recorded in the time domain as a function of D.

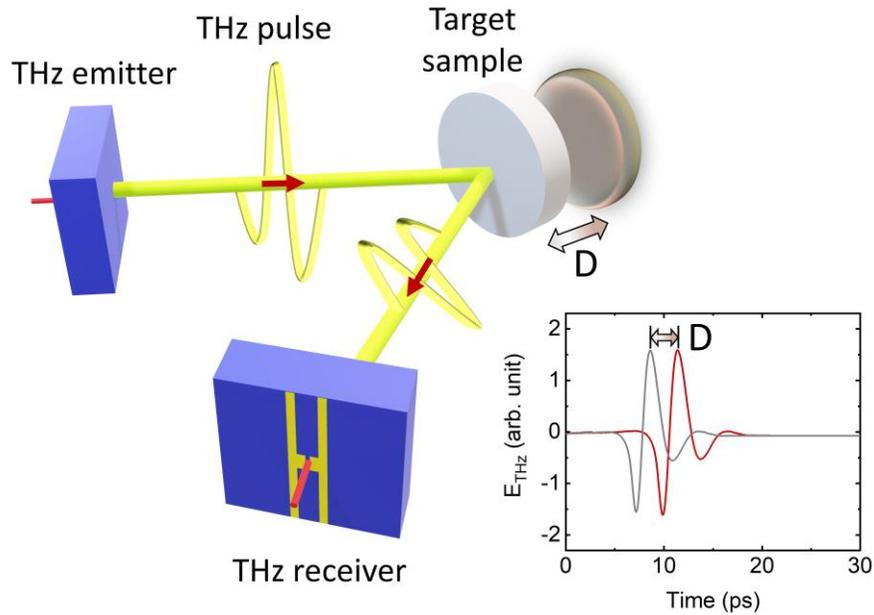

**Fig. 1.** Sketch of reflection type THz-TDS. Placement mismatch D is considered. Corresponding D on the time domain appears as the shifted THz pulse.

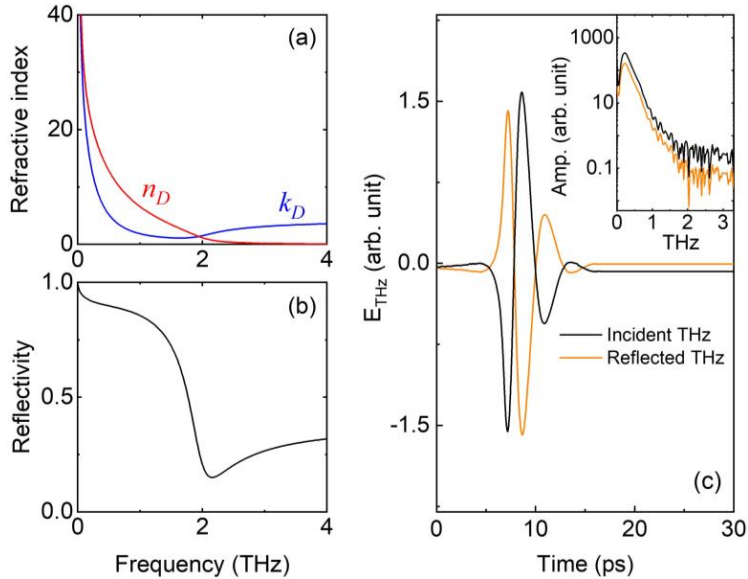

**Fig. 2.** (a) Metallic complex refractive index of $\tilde{n}_D=n_D+ik_D$ calculated with $\omega_p/2\pi=8.2$ THz, $\tau\approx3.7$ ps, and $\varepsilon_\infty=17$. (b) Reflectivity spectrum calculated from $\tilde{n}$ denoted in (a). (c) The THz pulse profile of incident $E_{inc}$ and computation result of the reflected $E_{rec}$. For the calculation of $E_{rec}$, Eq. (1) and $\tilde{n}$ presented in (a) are utilized. The inset shows the amplitude profiles of $E_{inc}$ and $E_{rec}$.

As for known $\tilde{n}=n+ik$, we first employed the Drude formalism which can representatively describe the optical properties determined by free carriers. Complex dielectric constant spectrum $\tilde{\varepsilon}_D=\varepsilon_{D1}+i\varepsilon_{D2}$ of Drude formalism is given by:[23]

$$\tilde{\varepsilon}_D=\varepsilon_\infty-\frac{\omega_p^2\tau}{\omega^2\tau-i\omega} \quad (2)$$

here $\tau$ is the relaxation time and $\omega_p$ is the plasma frequency, which is:

$$\omega_p = \sqrt{\frac{4\pi n_e e^2}{m^*}} \quad (3)$$

where $n_e$ and $m^*$ are the carrier density and the effective mass, respectively. In order to consider strong metallic features of $\tilde{n}$ in the THz frequency range, we refer to $\omega_p/2\pi=8.2$ THz, $2\pi\tau \approx 3.7$ ps, and $\varepsilon_\infty=17$. Finally, $\tilde{n}$ can be obtained by the relation of $\tilde{n} = \sqrt{\tilde{\varepsilon}}$. $\tilde{n}$ calculated with parameters above $\tilde{n}_D = n_D + ik_D$ is shown in Fig. 2(a) and its corresponding reflectivity spectrum is presented in Fig. 2(b). The effective plasma frequency is defined by $\omega_p/\sqrt{\varepsilon_\infty}$, which is about 2 THz. This effective plasma frequency corresponds to the cutoff frequency for the external electric fields, suggesting that one can observe the lowest reflectivity at the effective plasma frequency and higher reflectivity upon decreasing frequency from the effective plasma frequency. In other words, most of external electromagnetic waves are reflected by the plasma oscillation when the frequency is below $\omega_p/\sqrt{\varepsilon_\infty}$. The incident $E_{inc}$ and calculation result of the reflected THz beam $E_{rec}$ are presented in Fig. 2(c). As can be seen, $E_{rec}$ is inverted compared to $E_{inc}$. This fact is attributed to the fixed-end reflection where the reflector has a higher value of $\tilde{n}$ than the vacuum. The corresponding amplitude spectra for $E_{inc}$ and $E_{rec}$ are illustrated in the inset of Fig. 2(c), obtained by the numerical Fourier transform. One can confirm that $E_{rec}$ is slightly reshaped compared to $E_{inc}$; both amplitude spectra below 0.5 THz are almost identical whereas the amplitude spectrum for the sample becomes a lower curve as increasing frequency from 0.5 THz. This can be understood by referring to the reflectivity spectrum shown in Fig. 2(b).

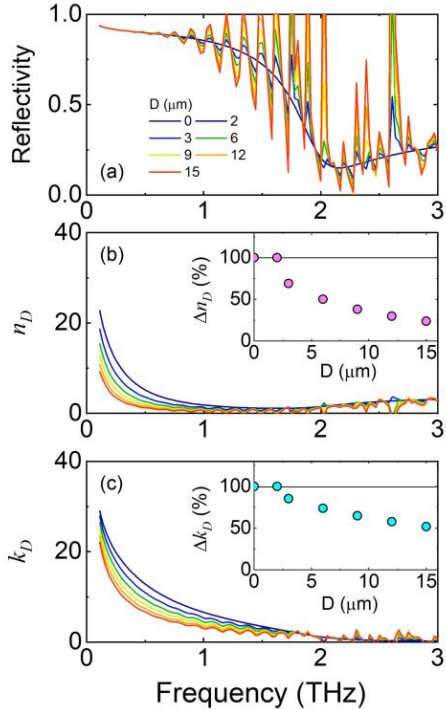

**Fig. 3.** Displacement D-dependent reflectivity (a), real (b), and imaginary (c) parts of complex refractive index. The insets in (b) and (c) show the accuracy of corresponding values at 0.5 THz.

Using Eqs (1)-(3), we can calculate reflected THz pulse $E_{rec}$ under the placement mismatch with a certain displacement D between the sample and the reference. This means that D-dependent $\tilde{n}=n+ik$ can be investigated by designing optical systems based on the Fresnel equation, as:[24-26]

$$\mathcal{F}(E_{rec}) = \mathcal{F}(E_{inc}) \cdot r_{ab} \tag{4}$$

Here, $r_{ab}$ is the Fresnel coefficient for the reflection, which is given by:

$$r_{ab} = \frac{\tilde{n}_a - \tilde{n}_b}{\tilde{n}_a + \tilde{n}_b} \tag{5}$$

The subscripts a and b indicate the first and second mediums that the THz pulse encounters during propagation. In this simulation environment, a is air and b is the metallic medium with $\tilde{n}_D$. Note that the THz pulse is incident normal to the sample. By solving Eq. (4) with $E_{rec}$ obtained from various sets of D's, $\tilde{n}_D$ can be extracted as a function of D, and its real part $n_D$ and imaginary part $k_D$ of $\tilde{n}_D$ are presented in Fig. 3(b) and (c), respectively. The selected D is denoted in Fig. 3(a). We can confirm that both extracted $n_D$ and $k_D$ are identical to the employed $\tilde{n}_D$ when D is less than 3 μm. However, uncertainty significantly increases upon increasing D beyond 3 μm. For example, the accuracy of $n_D$ ($\Delta n_D$) and $k_D$ ($\Delta k_D$) at 0.5 THz and D=6 μm are about 50 % and 75 %, respectively (see the insets of Fig. 3(b) and (c)). When D increases further, the uncertainty gradually becomes larger. This fact suggests that the placement mismatch between the sample and the reference should be below 3 μm in order to obtain the reliable $\tilde{n}$. Figure 3(a) shows D-dependent reflectivity spectra, from which a notable feature, i.e., strong oscillatory behavior, can be confirmed and it becomes stronger as D increases. Interestingly, the overall reflectivity profile does not appear to be significantly affected by D. Therefore, one idea to find the best alignment condition experimentally is to monitor the oscillatory behavior in the reflectivity spectrum and to find the location showing the minimal oscillation. By using this alignment method, we can minimize the placement mismatch between the sample and reference material (less than D of 3 μm) thereby increasing the accuracy of the extracted $\tilde{n}$.

From now on, we examine the accuracy of $\tilde{n}=n+ik$ when the collective excitations, e.g., phonon, are located in the THz frequency range. To reproduce such collective excitations, we employed the single Lorentz function, and its dielectric constant $\tilde{\varepsilon}_L = \varepsilon_{L1} + i\varepsilon_{L2}$ is given by:[27-29]

$$\tilde{\varepsilon}_L = \varepsilon_\infty + \frac{\omega_p^2}{(\omega_0^2 - \omega^2) - i\omega/\tau} \tag{6}$$

Here, $\omega_0$ is the center frequency of the oscillation, set to 1.5 THz. $\tau$ determines the width of the Lorentzian profile, and we employed $2\pi\tau = 2$ THz. The corresponding $\tilde{n}_L = n_L + ik_L$ is provided in the inset of Fig. 4(a). Comparing $\tilde{n}_L$ to $\tilde{n}_D$, $\tilde{n}_L$ generally has a higher value than $\tilde{n}_D$ when above 1 THz. The frequency component of the incident THz pulse $E_{inc}$ is mostly confined to below 1 THz (see the inset of Fig. 2(c)). This suggests that a higher reflected THz pulse $E_{rec}$ is expected with $\tilde{n}_D$, which can be confirmed by the direct comparison of $E_{rec}$ recorded in the time domain (see Fig. 2(c) and Fig. 4(a)).

The dependence of $\tilde{n}_L$ on displacement D is shown in Fig. 4(c) and (d). As can be seen, the accuracy of $n_L$ and $k_L$ values significantly decreases when D exceeds 3 μm, and their profiles deviate considerably from their original profile at D=0 μm. This observation is more clearly illustrated in the insets of Figs. 4(c) and (d), which display the change of $n_L$ and $k_L$ at $\omega_0$ of 1.5 THz as a function of D. Our calculations indicate that the accuracy of extracted $\tilde{n}$ is crucial for the placement mismatch between the sample and the reference, especially when the sample has the collective excitations in the THz frequency range. Interestingly, the reflectivity obtained with the extracted $\tilde{n}_L$ shows the comparable profile as when D=0 μm, except for the strong oscillatory behavior. Namely, the reflectivity profile does not appear to strongly depend on D, which is the similar observation for the case of $\tilde{n}_D$. This means that the optimal optical alignment condition for D can be determined by gradually translating D with the resolution of less than 1 μm and identifying the D that exhibits the minimum oscillatory behavior.

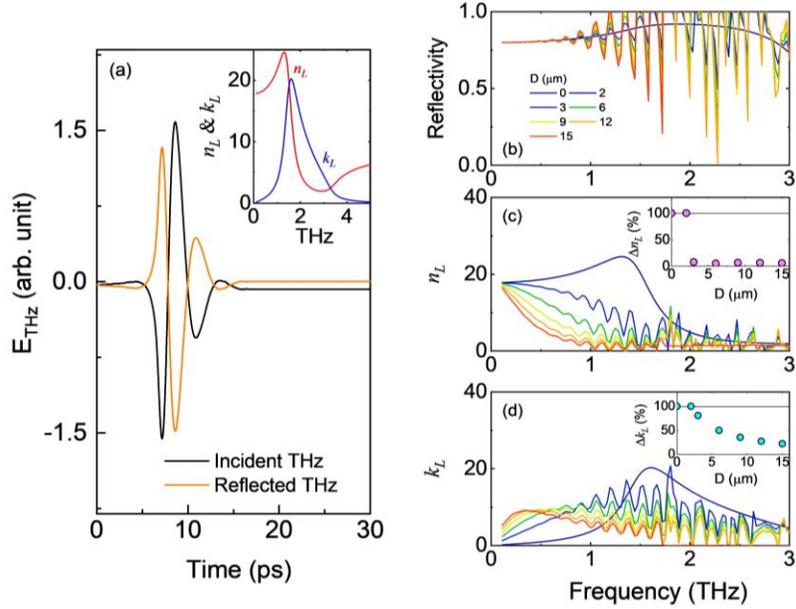

**Fig. 4.** (a) Incident THz and reflected THz pulses recorded in the time domain. The inset shows the complex refractive index obtained from the Lorentz model. Displacement D-dependent reflectivity (b), real (c), and imaginary (d) parts of complex refractive index. The insets in (c) and (d) show the accuracy of corresponding values at 1.5 THz.

## 3. CONCLUSION

In conclusion, we quantitatively analyzed the accuracy of the extracted complex refractive index under the placement mismatch D between the sample and the reference. To examine accuracy, we employed two representative systems whose optical properties are primarily determined by free carriers and bound carriers. In both cases, we confirmed that the accuracy of the extracted complex refractive index decreases significantly when D exceeds 3 μm. Interestingly, the overall profile of the reflective spectrum does not seem to be considerably influenced by D. However, strong oscillatory behavior was observed when D deviates from the ideal condition. Thus, this strong oscillatory behavior can serve as the beacon for the alignment condition, suggesting that the best alignment condition for D can be found by translating D within a 1 μm scale to find the D manifesting the minimized oscillatory features. We believe that our quantitative analysis can provide guidelines for spectroscopic investigations by extracting the optical responses functions based on the reflection type of THz-time-domain spectroscopy.